\documentclass{midl} 

\usepackage{mwe} 
\usepackage{textcomp}
\usepackage{xfrac}

\jmlrproceedings{MIDL}{Medical Imaging with Deep Learning}
\jmlrpages{}
\jmlryear{2019}
\jmlrworkshop{MIDL 2019 -- Extended Abstract Track}

\title[CT EFoV Using Channels Extension and Deep Learning]{CT Field of View Extension Using 
Combined Channels Extension and Deep Learning Methods}

\midlauthor{\Name{{\'E}ric Fourni{\'e}} \Email{eric.fournie@siemens-healthineers.com}\\
  \Name{Matthias {Baer-Beck}} \Email{matthias.baer@siemens-healthineers.com}\\
  \Name{Karl Stierstorfer} \Email{karl.stierstorfer@siemens-healthineers.com}\\
  \addr Siemens Healthcare GmbH,
  Computed Tomography,
  Siemensstr. 3,
  91301 Forchheim, Germany}

\begin{document}

\maketitle

\begin{abstract}
  This paper proposes a method to extend the field of view of computed 
  tomography images. In a first step, the field of view is increased by 
  extrapolating linearly the outer channels in the sinogram space. The 
  modified sinogram is then used to reconstruct extended field of view (EFoV) images 
  containing artifacts due to the channels extension. In a second step, those artifacts 
  are reduced by a deep learning network in image space. \\
  The proposed method has been evaluated on a collection of clinical scans. The resulting 
  volumes have been checked for consistency and plausibility and compared to an existing 
  state of the art EFoV method. \\
\end{abstract}


\section{Introduction}

\begin{figure}[ht]
  \begin{center}
  \includegraphics[width=0.35\columnwidth]{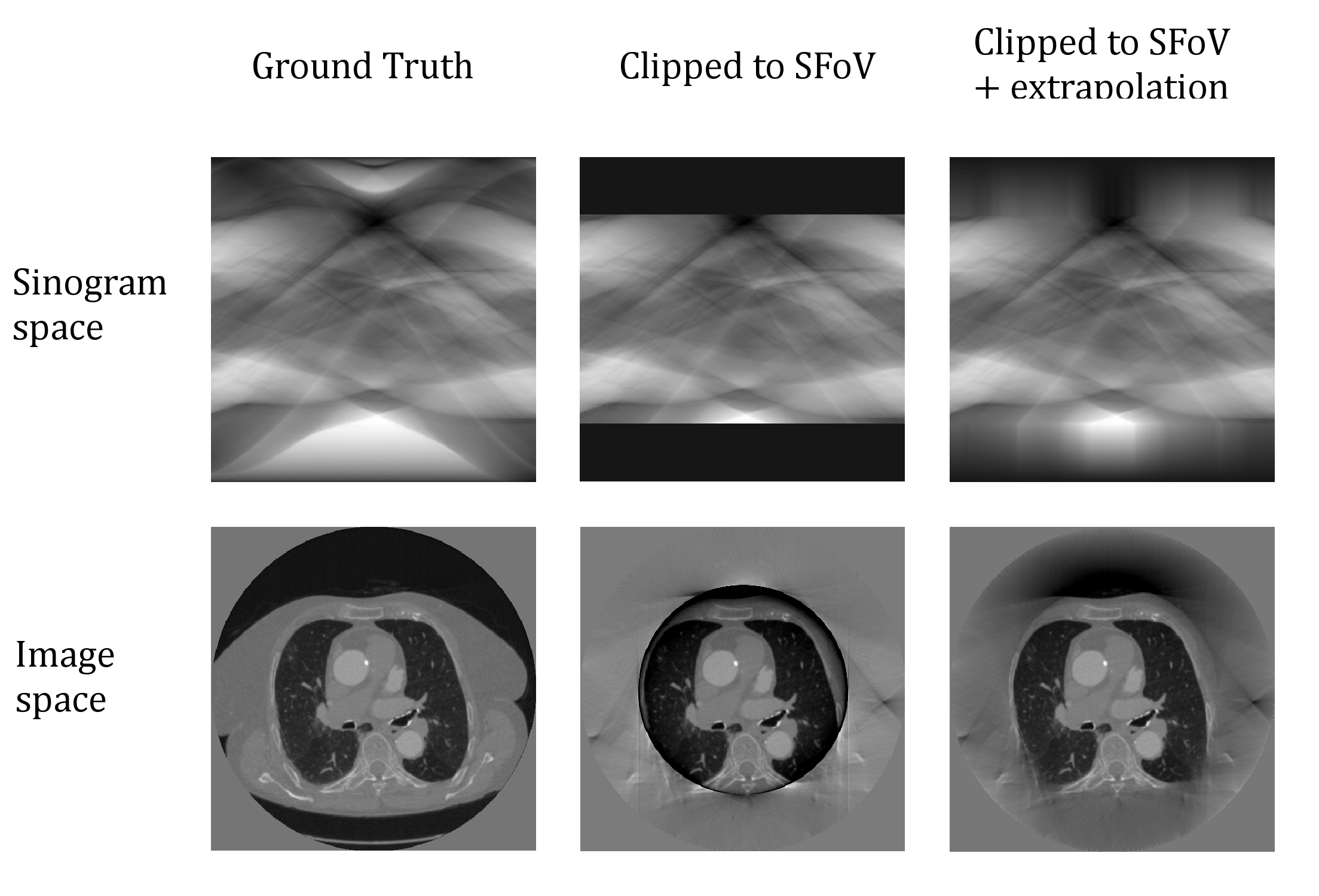}
  \caption{Field of view extension by channel padding or extrapolation. 
  \label{ch_extension} 
   }
   \end{center}
\end{figure}

X-ray computed tomography (CT) reconstructs slices of the scanned structure 
from the sinograms acquired by rotating a detector array and an X-ray emitting 
tube around the scanner gantry. One of its fundamental limitation is that only 
data that is located inside the scan field of view (SFoV) of the CT scanner 
can be reconstructed without artifacts. The size of the SFoV is determined by the 
geometrical properties of the CT scanner, i.e. by the size of the detector and by the 
distance of the X-ray tube and the detector from the isocenter of the CT scanner.
Sinograms nonetheless contains information from regions outside of the SFoV. Artificially 
extending their angular width, for example by padding or extrapolating 
the channels as in fig. \ref{ch_extension}, will allow for a partially 
reconstructed EFoV region polluted by artifacts. 

The proposed method first extends the sinograms in channel direction, then uses a U-Net 
\cite{Ro15} to clean up the artifacts. The result 
is a realistic and computationally cheap extension of the field of view. Compared to 
existing methods \cite{Hs04, Ku07,Br08}, the proposed approach delivers improved results at a 
low computational cost.

\section{Method}

The proposed method consists of a two-step approach. Let us consider a scanned object 
\(y\) and its sinogram obtained by the Radon transform \(\Re(y)\). 
In a first step, the angular width of the sinogram is increased by 
linear extrapolation \(g(\Re(y))\) of the outermost channels linearly towards zero, 
thus increasing the number of channels by a factor corresponding to \(EFoV/SFoV\).
We can now use any method \(\Re^+\) such as filtered backprojection (FBP) to 
reconstruct an EFoV image \(X=\Re^+(g(\Re(y)))\) containing artifacts. The SFoV 
region is only slightly influenced by the extrapolation via the convolution step of 
the FBP and therefore can be used for diagnostic purposes without limitation.

The artifacts produced outside of the SFoV region in \(X\) are then cleaned up by a 
U-Net \(F\) trained to match \((X,y)\) pairs. The output of the network \(\hat{y}=F(X)\) 
can then be used as a realistic estimation of \(y\) for the EFoV region. With this 
approach, the problem faced is the reduction of artifacts instead of inpainting 
large portions of the image, thus largely reducing the space of acceptable solutions.

\begin{figure}[ht]
  \begin{center}
    \includegraphics[width=0.6\columnwidth]{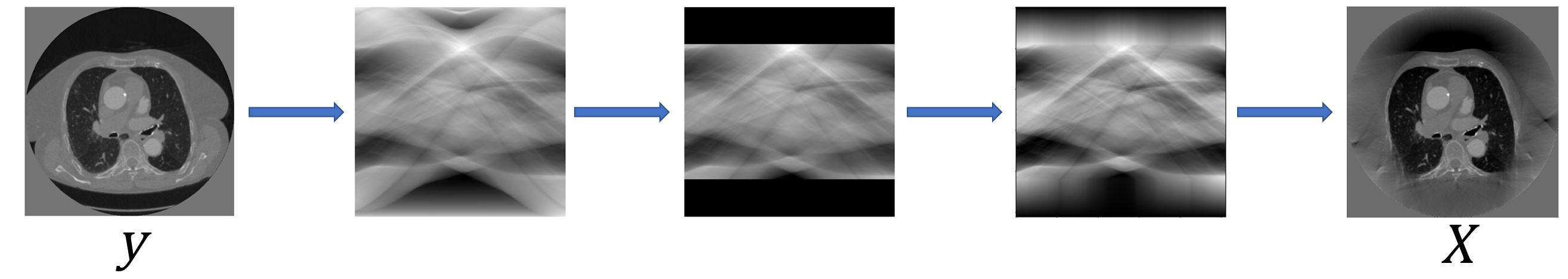}
    \includegraphics[width=.6\columnwidth]{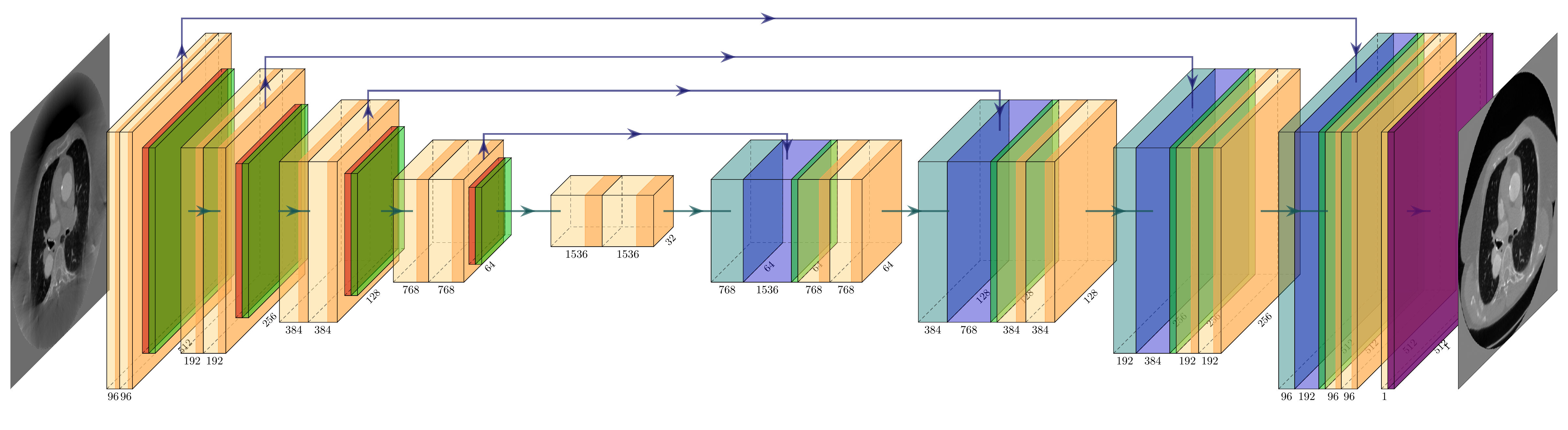}
  \includegraphics[width=.20\columnwidth]{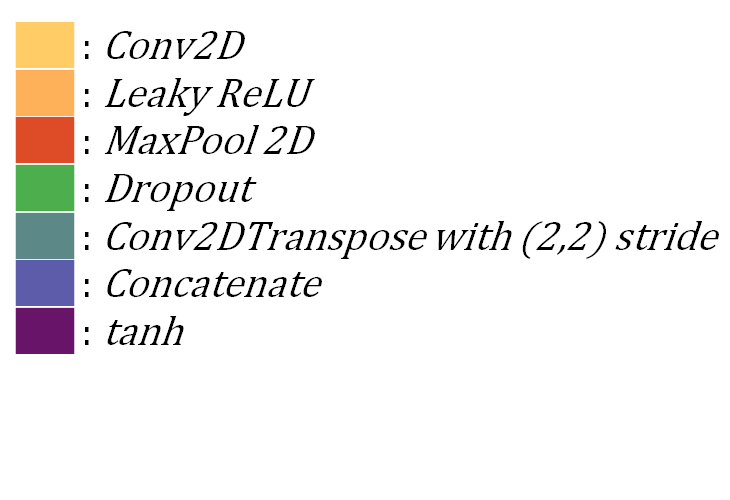}
  \caption{Data generation pipeline and network architecture
  \label{deep_EFoV} 
   }
   \end{center}
\end{figure}

To remediate possible modifications of the SFoV region by the network, it is replaced 
by the corresponding SFoV of \(X\). In order to improve the final 
SFoV/EFoV transition, the network output is adapted to the input distribution:

\[\hat{y}=\frac{\hat{y}-\mu_{\hat{y}}}{\sigma_{\hat{y}}}\cdot\sigma_X+\mu_X\]

To train the network, a collection of clinical CT scan slices \(y\) are projected 
to their Radon transform to obtain sinograms. The sinograms are extended by extrapolation 
of the outer channels as described previously (cf. fig. \ref{ch_extension} and 
\ref{deep_EFoV}) and backprojected to images \(X\) used as inputs of the network. The 
loss function is only computed in the EFoV region and is defined as the weighted sum of 
the structural dissimilarity \(DSSIM=(1-SSIM)/2\) \cite{Wa04} and the mean squared error.

\section{Results and Conclusions}

To evaluate performance on clinical data, a plugin implementing the various steps 
of the method including execution of the trained network has been developed for ReconCT, 
a proprietary offline reconstruction software developed by Siemens Healthineers.
ReconCT can also reconstruct EFoV images using the existing state of the art HDFoV 
method. 

\begin{figure}[ht]
  \begin{center}
    \includegraphics[width=1\columnwidth]{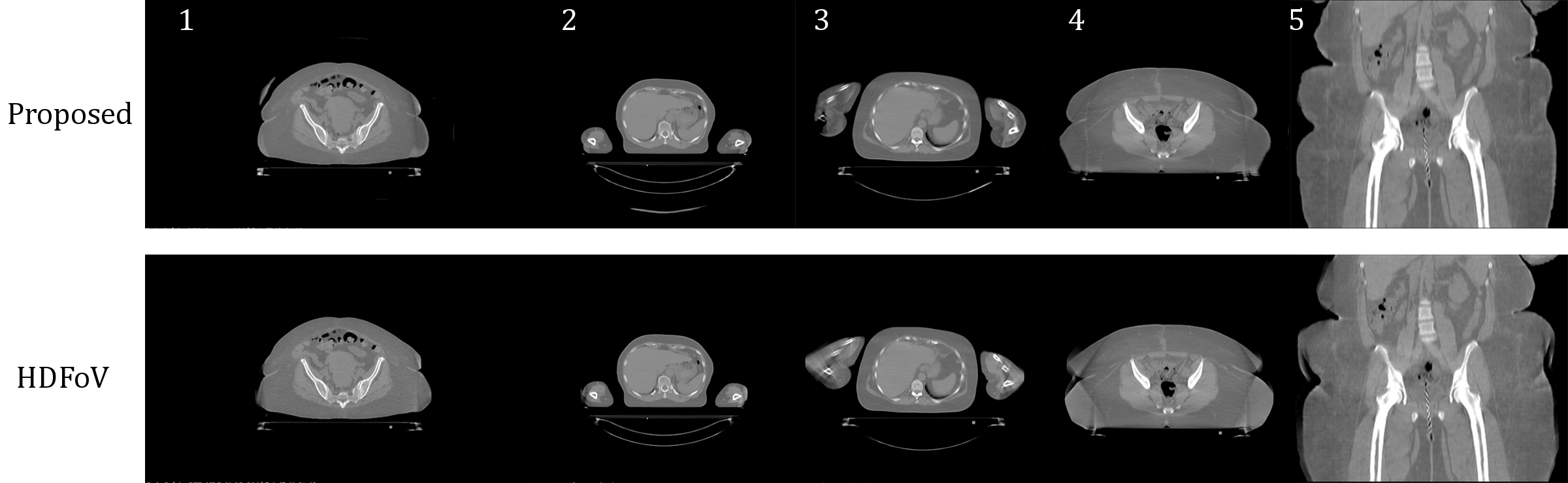}
  \caption{Comparison between our method (top) and HDFoV (bottom)
  \label{comparison} 
   }
   \end{center}
\end{figure}

Scanner raw data corresponding to 11 patients and 4 phantoms has been collected and 
EFoV images reconstructed using the proposed method and HDFoV. While no ground truth 
exists for the patient data sets, reconstructions have been checked for artifacts errors 
and plausibility in axial, coronal and sagittal slices. Some qualitative results are 
presented in fig. \ref{comparison}. When comparing the results depicted in figure 3 
it shows up that the results with the proposed method are superior to the HDFoV 
results for cases 2, 4 and 5. For cases 1 and 3 the proposed method cannot outperform 
the HDFoV method as it either introduces artificial structures (case 1) or fails to 
reconstruct some parts of the patient anatomy (arm in case 3).

Extrapolation by linear extrapolation followed by the removal 
of artifacts in the corresponding reconstruction using a deep learning network proved to be 
an efficient method to extend the field of view of CT scans and shows potential to 
improve quality of EFoV reconstructions. 
Like all methods aiming for the reconstruction of volume areas 
using incomplete projection data, the EFoV regions cannot be used for diagnostic 
but they are sufficient for an array of clinical applications where the normal SFoV is too 
narrow, such as radio therapy planning or the imaging of obese patients.


\bibliography{fournie19}

\end{document}